\documentclass[%
 aip,
rsi,%
 amsmath,amssymb,
 reprint,%
]{revtex4-1}

\usepackage[dvipsnames]{xcolor}
\usepackage{graphicx}
\usepackage{dcolumn}
\usepackage{bm}

\usepackage[ruled,vlined]{algorithm2e}

\usepackage{algpseudocode} 
\usepackage{dcolumn}
\usepackage{epstopdf}
\usepackage{xcolor}
\usepackage{natbib}
\usepackage{subcaption} 
\usepackage{float}
\usepackage{tabularx}
\usepackage{mathrsfs}
\usepackage{upgreek}
\usepackage[a]{esvect}
\setcitestyle{square}
\usepackage[toc,page]{appendix}
\usepackage[font={small,normalfont}]{caption}
\usepackage[colorlinks=true, allcolors=blue]{hyperref}
\newcommand{\rc}[1]{\textcolor{black}{#1}}

\newcommand{\gc}[1]{\textcolor{black}{#1}}

\begin{document}
\title{Particle integrator for particle-in-cell simulations\\of ultra-high intensity laser-plasma interactions}

\author{Kavin Tangtartharakul}
\affiliation{Department of Mechanical and Aerospace Engineering, University of California San Diego, La Jolla, CA 92093, USA}

\author{Guangye Chen}
\affiliation{Los Alamos National Laboratory, Los Alamos, New Mexico 87545, USA}

\author{Alexey Arefiev}
\affiliation{Department of Mechanical and Aerospace Engineering, University of California San Diego, La Jolla, CA 92093, USA}

\date{\today}
\begin{abstract}
Particle-in-cell codes are the most widely used simulation tools for kinetic studies of ultra-intense laser-plasma interactions. Using the motion of a single electron in a plane electromagnetic wave as a benchmark problem, we show surprising deterioration of the numerical accuracy of the PIC algorithm with increasing normalized wave amplitude for typical time-step and grid sizes. Two significant sources of errors are identified: strong acceleration near stopping points and the temporal field interpolation. We propose adaptive electron sub-cycling coupled with a third order temporal interpolation of the magnetic field and electric field as an efficient remedy that dramatically improves the accuracy of the particle integrator.

\end{abstract}

\maketitle
\section{Introduction} \label{Sec:introduction}

The ELI-NP laser facility~[\onlinecite{PW_Lasers_2019}] has recently demonstrated that their laser system is able to achieve the projected laser power of 10~PW~[\onlinecite{lureau_matras_chalus_derycke_morbieu_radier_casagrande_laux_ricaud_rey_et_all_2020}]. Even based on conservative estimates, such a powerful laser pulse will be able to achieve intensities exceeding $5 \times 10^{22}$~watt/cm$^2$. The corresponding normalized laser amplitude 
\begin{equation}
    a_0 \equiv |e| E_0 / m c \omega
\end{equation}
would exceed {$a_0 \approx 150$}, where $E_0$ is the peak amplitude of the laser electric field, $\omega$ is the laser frequency, $c$ is the vacuum speed of light, and $m$ and $e$ are the electron mass and charge. We define $\omega = 2 \pi c / \lambda$, where $\lambda = 820$~nm is the vacuum wavelength of the laser. The significance of reaching such high values of $a_0$ is that the motion of laser-irradiated electrons inside a target would become ultra-relativistic. 

Simulations of laser-matter interactions at $a_0 > 1$ typically require a kinetic approach. The particle-in-cell (PIC) approach is one such approach that has been extensively used by the community to gain valuable physics insights. A standard PIC algorithm consists of two main modules: a particle pusher and a field solver, which commonly employ the Boris algorithm~[\onlinecite{birdsall1985plasma}] and the Yee scheme~[\onlinecite{yee1966numerical}], respectively.

Several recent publications showed that applying the conventional PIC algorithm at $a_0 \gg 1$ might be problematic at least in certain regimes~[\onlinecite{Arefiev_PoP_2015-resolution,PushingParticlesExtremeFieldsGordon2017,crossbeam2019}]. A specific regime of interest for this study is the one where a laser-irradiated electron is able to travel a longitudinal distance that significantly exceeds the laser wavelength $\lambda$. In this regime, the electron experiences alternating acceleration and deceleration while periodically coming to a complete stop during its predominantly forward motion. Even for given fields with an analytical expression, the Boris algorithm fails to recover the particle dynamics near the stopping points if $a_0$ is increased for a fixed time-step $\Delta t$~[\onlinecite{Arefiev_PoP_2015-resolution}]. This leads to significant errors along the rest of the electron trajectory and impacts the maximum energy gain. 

One needs to dramatically reduce the time-step $\Delta t$ in a PIC algorithm with $a_0 \gg 1$ to accurately reproduce the electron energy gain in the described regime. This is usually achieved by reducing the cell size. For instance, in a 1D simulation, we keep the ratio $\Delta x / c \Delta t$ close to unity, where $\Delta x$ is the cell size. Otherwise, the numerical dispersion errors for electromagnetic waves representing the laser become significant and this can adversely affect the electron acceleration as well. For example, 80 cells per wavelength were used to correctly reproduce electron acceleration at $a_0 \approx 8.5$ in an extended relativistically transparent preplasma~[\onlinecite{Sorokovikova_2016}].

\begin{figure*}
    \centering
    \includegraphics[width=\textwidth]{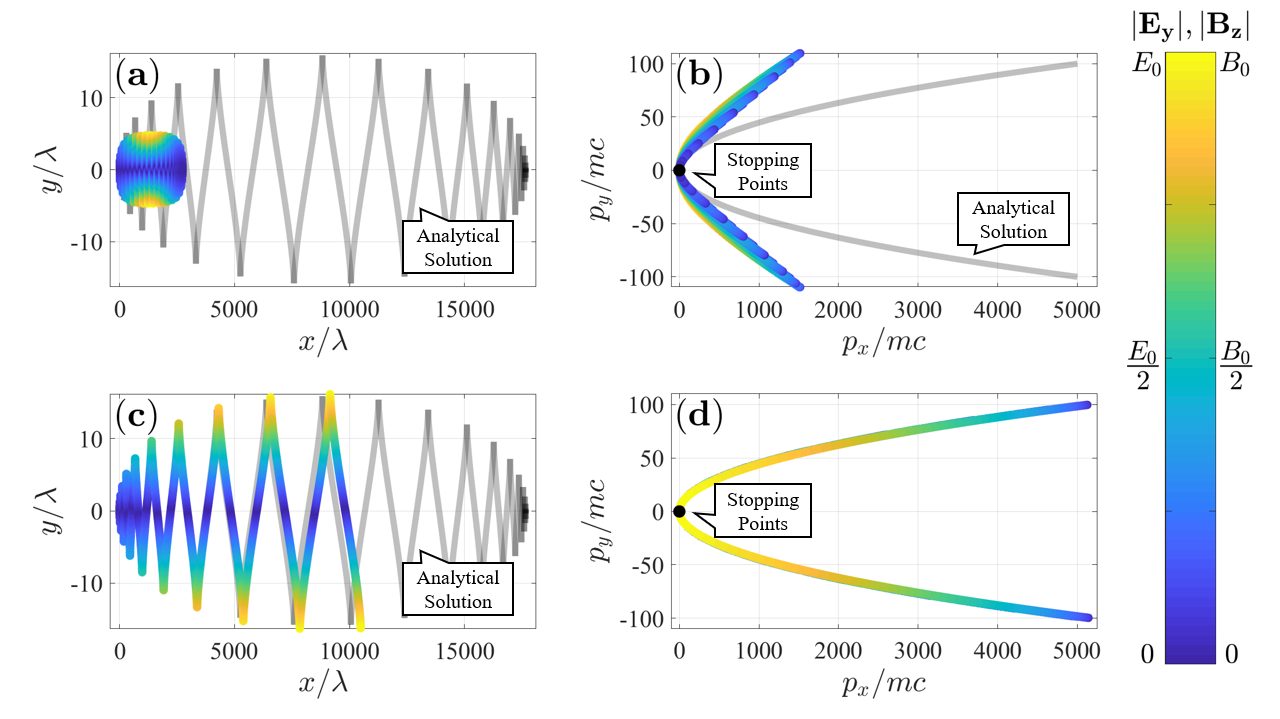}
    \caption{Trajectories of an electron irradiated by a wave with $a_0 = 100$ (left column) and the corresponding momentum (right column). The gray scale curves are the analytical solution. The color-coded curves are results from perfectly propagating PIC algorithm with particle sub-cycling ($\lambda / \Delta t c = T/\Delta t = 75$, \gc{$\lambda / \Delta x \approx 74.26$}, \gc{$c \Delta t / \Delta x  = 1 / 1.01=0.99$}, and $\Psi_{max}=0.01$): panels (a) and (b) are obtained using $1^{st}$ order temporal interpolation of the fields; panels (c) and (d) are obtained using $3^{rd}$ order temporal interpolation of the fields. The black point represents the stopping point. The color gradient represents the magnitude of the $E$ and $B$ fields.}
    \label{momentum}
\end{figure*}

\begin{figure*}
    \centering
    \includegraphics[width=0.95\columnwidth]{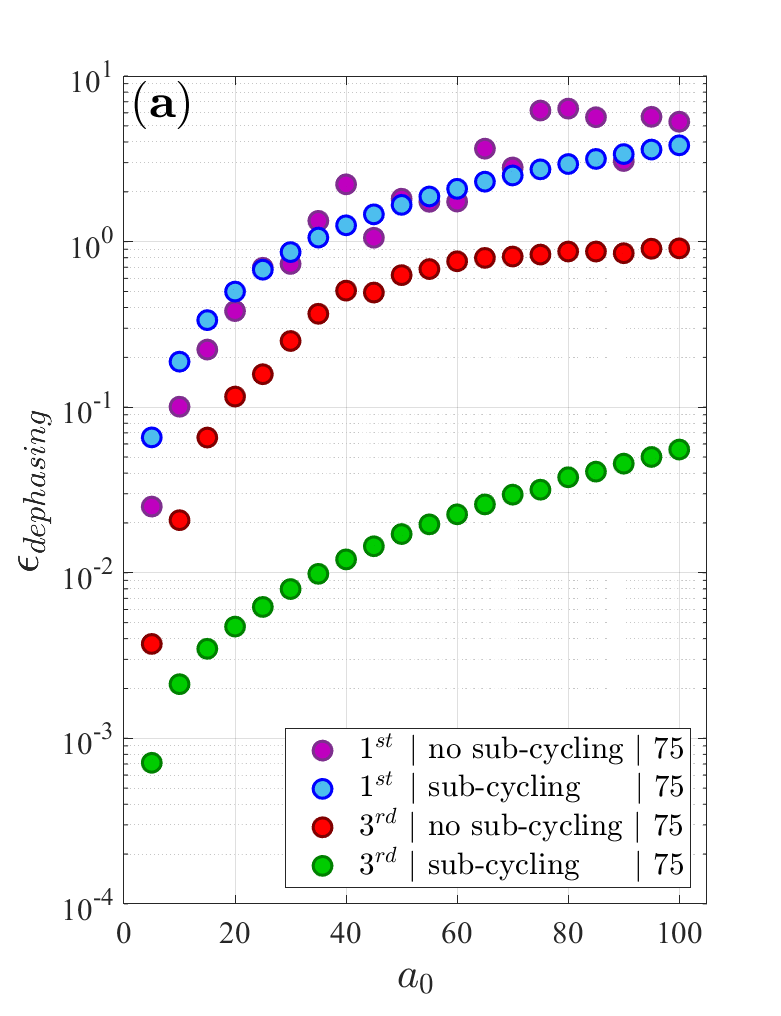}\hspace{1cm}\includegraphics[width=0.95\columnwidth]{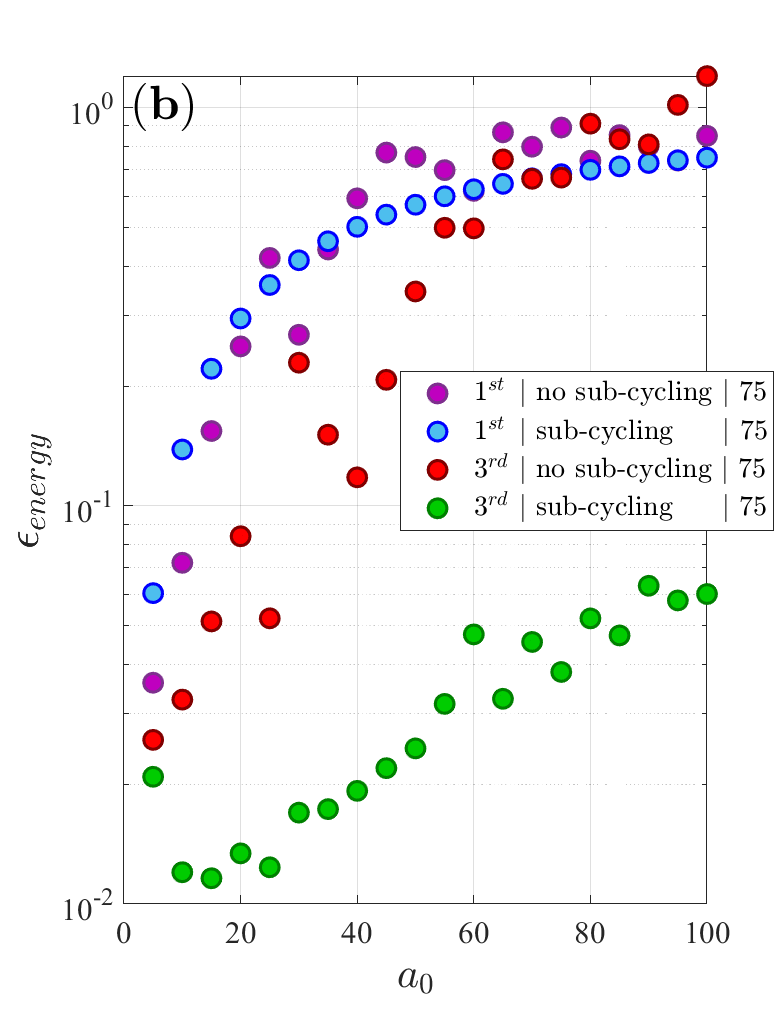}
    \caption{Relative dephasing, $\epsilon_{dephasing}$, and energy, $\epsilon_{energy}$, errors for different versions of the PIC algorithm over a range of laser amplitudes $a_0$. Panel (a) is obtained using the standard PIC with and without the sub-cycling. Panel (b) is obtained using the perfectly propagating PIC with and without the sub-cycling. The data sets shown with red and green markers in both panels are obtained by replacing the $1^{st}$ order interpolation with the $3^{rd}$ order temporal interpolation for the fields detailed in Sec.~\ref{Sec-6}. All simulations use $\Delta t = \lambda / 75 c$ \gc{and $\Delta x \approx \lambda / 74.26$ to satisfy $c \Delta t / \Delta x  = 0.99$}.}
    \label{fig:compare}
\end{figure*}

In this paper, we describe a particle integrator that can overcome the described difficulty and accurately reproduce electron acceleration at $a_0 \gg 1$ without having to use the number of time-steps that increases proportionally to $a_0$~[\onlinecite{Arefiev_PoP_2015-resolution}]. Two key components are used in combination to improve the accuracy: 1) adaptive temporal sub-cycling of the conventional particle pusher employing the Boris algorithm~[\onlinecite{birdsall1985plasma}]; 2) third-order (or higher), rather than the conventional first-order, temporal interpolation of electric and magnetic fields used by the pusher. Our key finding is that the Boris algorithm using sub-cycling~[\onlinecite{Arefiev_PoP_2015-resolution}] may perform poorly when implemented into a standard PIC algorithm. The electromagnetic fields must be temporally interpolated with high-order polynomials for the sub-cycling to yield significant benefits. When using appropriately high-order temporal field interpolations, the sub-cycling improves simulation performance (see Fig.~\ref{fig:compare}) at $a_0 \gg 1$ by at least an order of magnitude without increasing significantly the computational cost of the simulation. \rc{Even though we consider a specific implementation that involves the sub-cycling, these findings regarding the need for the high-order interpolation apply to other particle pushers designed to improve the performance at $a_0 \gg 1$~[\onlinecite{PushingParticlesExtremeFieldsGordon2017, GORDON2021_SpecialUnity}].}

This paper is organized as follows: In Section~\ref{Sec:model}, we introduce the model problem and its key analytical results. In Section~\ref{Sec-extra}, we review how the conventional PIC algorithm is temporally discretized and introduce a ``perfectly propagating PIC'' algorithm that is then used to clearly identify the impact of temporal interpolation. In Section~\ref{Sec-4}, we show that implementing particle sub-cycling into a PIC algorithm that utilizes linear temporal interpolation is not sufficient to dramatically improve the simulation. We then explain, in Section~\ref{Sec-5}, why the linear interpolation is responsible for the failure of the sub-cycling algorithm. In Section~\ref{Sec-6}, we describe a higher order temporal interpolation procedure for the fields and show that the third order interpolation dramatically improves the results for the model problem when used together with the sub-cycling. Finally, in Section~\ref{Sec-7}, we quantify the performance of the improved PIC algorithm by comparing it with that of a conventional PIC algorithm for the model problem. 


\section{Description of Model Problem} \label{Sec:model}

In this section we review the solution of a well-known problem where a single electron is irradiated in vacuum by a linearly polarized electromagnetic wave~[\onlinecite{shebalin1988exact}] . This problem provides an important test case for evaluating the performance of particle integrators at $a_0 \gg 1$ and it is used extensively in the remainder of the paper.

We consider an ultra-intense plane electromagnetic wave with a Gaussian temporal profile that propagates, in a vacuum, along the $x$-axis with the wave electric field directed along the $y$-axis. The field evolution in this case is fully described by a normalized vector potential that only has a $y$ component and that is only a function of the phase variable $\xi \propto t-x/c$. We denote this component as $a$, such that 
\begin{eqnarray}
    && E_y = B_z = -\frac{m \omega c} {|e|} \thinspace \frac{d a}{d \xi}, \label{laser5} \\
    && a(\xi) \equiv a_0 \thinspace \exp \big[ -(\xi - \xi_0)^2 / (2 \sigma^2) \big] \thinspace \sin( \xi ), \label{laser1}\\
    && \xi \equiv \omega (t - x / c), \label{laser3} 
\end{eqnarray}
where $t$ is the time and $a_0$ is the peak amplitude. In the specific examples that follow we set $\xi_0 = -160\pi$ and $\sigma = 8\pi$. Note that the phase velocity in this case is equal to the speed of light, $v_{ph} = c$. The group velocity is also equal to $c$, such that the pulse preserves its shape during propagation.

An analytical solution exists for an electron irradiated by the described wave. Appendix A provides details on how to obtain a general solution, so here we only summarize the key results. We take an electron that is at rest prior to the arrival of the wave, i.e. at $a = 0$. Then the electron momentum in the wave is given by
\begin{eqnarray}
    && p_x/m c = a^2(\xi)/2 \geq 0, \label{integral motion x}\\
    && p_y/m c = a(\xi). \label{eq:6}
\end{eqnarray}
The relativistic $\gamma$-factor defined as $\gamma = \sqrt{1 + p^2/ m^2 c^2}$ is
\begin{equation}
    \gamma = 1 + a^2(\xi)/2. \label{theoretical gamma}
\end{equation}
As the electron moves in the wave, the following quantity remains conserved
\begin{equation}
    R \equiv \gamma - p_x /mc = 1. \label{theoretical dephasing}
\end{equation}
It can be shown that
\begin{equation} \label{R-def}
    R = \frac{\gamma}{\omega} \frac{d \xi}{d t},
\end{equation}
so $R$ represents the electron dephasing rate in the electron's instantaneous rest frame~[\onlinecite{Arefiev_Enhancement_Laser_Channel_2014}].

An example of the described analytical solution in Eqs.~(\ref{integral motion x}) and (\ref{eq:6}) is shown by the gray curve in Fig.~\ref{momentum}b for $a_0 = 100$. The corresponding electron trajectory is shown by the gray curve in Fig.~\ref{momentum}a, where we set the electron initial location as $x = y = 0$. These figures illustrate that $p_x$ dominates the trajectory at $a_0 \gg 1$, which leads to a large displacement in the $x$ direction. In Fig.~\ref{momentum}, numerical solutions, that will be discussed in the next section, are plotted on top of the analytic solution with the color-coding indicating the strength of the electric, $|E_y|$, and magnetic, $|B_z|$, fields at the electron location. The electron reaches its most energetic state when the $E$ and $B$ fields vanish, which corresponds to $|a(\xi)|=a_0$. And when the field strength is at a local maximum, the electron reaches a motionless position. We call these positions the stopping points. We single out these points because the electron dynamics in their vicinity during numerical integration has strong impact on the subsequent electron acceleration.


\section{PIC algorithm and temporal discretization} \label{Sec-extra}

The focus of the paper is on simulations of electron acceleration by an ultra-high intensity laser pulse and the accuracy improvements that can be achieved within the framework of a PIC algorithm. The details of a standard PIC algorithm are well-known and can be found in Ref.~[\onlinecite{birdsall1985plasma}]. Here we review the key aspects of the temporal discretization with a fixed time-step $\Delta t$ to provide the necessary context for the discussion that follows. 

The model problem detailed in Sec.~\ref{Sec:model} can be simulated using a one-dimensional (1D) version of a standard PIC code. The difficulties mentioned in Sec.~\ref{Sec:introduction} can already be observed in 1D, so we limit our analysis to a 1D version of a PIC code in order to make the discussion more compact. 

In a standard PIC algorithm, there are two types of quantities: those known at $t = n \Delta t$ and denoted by $n$; and those known at $t = (n + 1/2) \Delta t$ and denoted by $n+1/2$, where $n$ is an integer (the simulation starts at $n = 0$). \gc{The field equations are discretized in space and time using the Yee scheme. In our problem, there are only $E_y$ and $B_z$ field components, so we have}
\begin{eqnarray}
\gc{E_y|_{n}^{i+1/2} = E_y|_{n-1}^{i+1/2} - \frac{c \Delta t }{\Delta x} \left( B_z|_{n-1/2}^{i+1}-B_z|_{n-1/2}^i \right)}, &&\label{eq:Ey} \\
\gc{B_z|_{n+1/2}^i = B_z|_{n-1/2}^i - \frac{c \Delta t }{\Delta x} \left( E_y|_n^{i+1/2}-E_y|_n^{i-1/2} \right)}. && \label{eq:Bz} 
\end{eqnarray}
\gc{Here the superscript represents the spatial discretization with a cell size $\Delta x$, such that $E_y|^{i+1/2}$ is the field at $x = x^{i + 1/2} \equiv (i + 1/2) \Delta x$ and $B_z|^{i}$ is the field at $x = x^{i} \equiv i \Delta x$.
The Boris particle pusher that we use to advance the considered electron requires the knowledge of $B_z$ at $t = n \Delta t$. The standard approach, which in this paper is called ``standard PIC algorithm'', uses a linear temporal interpolation:}
\begin{equation} \label{linear interp orig}
     \gc{B_z|_{n}^i = \frac{1}{2} \left( B_z|_{n + 1/2}^i + B_z|_{n - 1/2}^i \right).}
\end{equation}
\gc{A shape function $S$ is used to obtain the fields $E_y$ and $B_z$ at the electron position $x_n$ (we denote these fields with a superscript $p$):} 
\begin{eqnarray}
    && \gc{E_y^p |_n = \sum_i E_y|_{n}^{i+1/2} S(x^{i+1/2} - x_n),} \label{eq:Enp}\\
    && \gc{B_z^p |_n = \sum_i B_z|_{n}^{i} S(x^{i} - x_n),} \label{eq:Bnp}
\end{eqnarray}
\gc{In this paper, we employ a triangular or tent shape function~[\onlinecite{birdsall1985plasma}]).} \gc{The electron is advanced using the standard Boris particle pusher detailed below:} 
\begin{eqnarray}
\bm{p}_{n+1/2}&=& \bm{p}^+ + \frac{q {\bm{E}^p|_n} \Delta t}{2}, \label{eq:p} \\
&&{\bm{p^+}}={\bm{p^-}} + {\bm{p'}} \times {\bm{{s}}}, \\
&&{\bm{p}^-}={\bm{p}_{n-1/2}} + \frac{q {\bm{E}^p|_n} \Delta t}{2}, \label{eq:p2} \\ 
&&{\bm{p}'}={\bm{p}^-} + {\bm{p}^-} \times {\bm{\Psi}}, \\
&&{\bm{{s}}}=\frac{2 {\bm{\Psi}}}{1+\Psi^2}, \\
&&{\bm{\Psi}}=\frac{q {\bm{B}^p|_n} \Delta t}{2 \gamma_n m_e c}, \label{eq:p3}\\
\gamma_{n} &=& \sqrt{1+\left(\frac{p^-}{m_ec}\right)^2}, \label{eq:gamman} \\
\gamma_{n+1/2} &=& \sqrt{1+\left(\frac{p_{n+1/2}}{m_ec}\right)^2}, \label{eq:gamma} \\
{\bm{v}_{n+1/2}}&=&\frac{{\bm{p}_{n+1/2}}}{m_e \gamma_{n+1/2}}, \label{eq:v} \\
{\bm{r}_{n+1}}&=&{\bm{r}_{n}}+\Delta t({\bm{ v}_{n+1/2}}). \label{eq:x}
\end{eqnarray}


In the standard PIC algorithm, the temporal discretization of the fields can impact the electron dynamics in two distinct ways: through numerical dispersion and through errors introduced by the temporal interpolation. It is well-understood that the numerical dispersion of the Yee scheme alters the electron dephasing rate and gradually distorts the envelope of the considered pulse. Its negative effects can be greatly reduced by setting the ratio $\Delta x / c \Delta t$ close to unity in 1D simulations without dramatically increasing the resolution. In contrast to that, the errors introduced by linearly interpolating the fields can only be reduced by reducing the time-step $\Delta t$. 

In order to clearly identify the impact of the temporal interpolation on electron dynamics, we introduce a reduced version of the PIC algorithm where the fields of the laser propagate according to the analytical expressions given by Eqs.~(\ref{laser5})~-~(\ref{laser3}), using the same $\Delta t$ and $\Delta x$ as the standard PIC algorithm. As a result, it does not solve the Maxwell's equations,  but retains the need for the temporal and spatial interpolations of the field from grid points to particles. In what follows, we refer to this algorithm as the ``perfectly propagating PIC''. The algorithm is particularly useful for simulating the described model problem because only the fields of the laser need to be accounted for in this case and this is exactly what is done in the reduced algorithm.


\section{Performance of PIC algorithms with first-order temporal interpolation} \label{Sec-4}

In this section, we examine how the standard and the perfectly propagating 1D PIC algorithms perform with the increase of $a_0$ when employing the first-order temporal field interpolation. We also introduce the concept of temporal sub-cycling and evaluate its impact on the performance of both algorithms.

A robust metric for evaluating the performance of a PIC algorithm is the maximum relative error in the desphasing rate, 
\begin{eqnarray}
    &&\epsilon_{dephasing} \equiv \left| \frac{R -R_{th}}{R_{th}} \right|_{max} = \left| R - 1 \right|_{max}.  \label{R error} 
\end{eqnarray}
Here $R_{th} = 1$ is the analytical result for the model problem and
\begin{equation}
    R = \gamma - p_x /mc
\end{equation}
is the dephasing calculated using the numerical values of the electron momentum. The advantage of using $\epsilon_{dephasing}$ is that it is straightforward to compute regardless of the numerical algorithm.

Fig.~\ref{fig:compare}a shows that $\epsilon_{dephasing}$ increases almost monotonically with the increase of $a_0$ in a simulation that uses the standard 1D PIC algorithm (purple points). \gc{The time-step is set at $\Delta t = T/75$, where $T = 2 \pi /\omega$ is the period of the considered wave.} We keep the ratio \gc{$c \Delta t / \Delta x  = 0.99$} fixed as we vary $a_0$ to ensure that the numerical dispersion remains unaffected by the scan. \gc{Accordingly, the grid size is set set to $\Delta x \approx \lambda/74.26$}. \gc{This resolution is similar to that used in Ref.~[\onlinecite{Sorokovikova_2016}] to correctly reproduce electron acceleration at $a_0 \approx 8.5$.} At $a_0 > 40$, the errors become \gc{more} significant, leading to strong deviations from the analytical trajectory shown by the gray curve in Fig.~\ref{momentum}a. As expected, the energy gain from the wave is also greatly impacted, with the results of a similar scan available in Ref.~[\onlinecite{Arefiev_PoP_2015-resolution}].

We then repeated the same scan using the perfectly propagating PIC algorithm and found that the results for $\epsilon_{dephasing}$ are very similar to those shown in Fig.~\ref{fig:compare}a. We have additionally examined how the energy gain deviates from the analytical solution along the electron trajectory. The corresponding metric is the maximum relative error in the energy gain defined as
\begin{eqnarray}
    &&\epsilon_{energy} \equiv \left| \frac{\gamma - \gamma_{th}}{\gamma_{th}} \right|_{max}, \label{gamma error}
\end{eqnarray}
where $\gamma_{th}$ is the analytical result given by Eq.~(\ref{theoretical gamma}). This comparison is possible due to the fact that the perfectly propagating PIC uses analytically prescribed fields and the theoretical value of $\gamma$ is known for these fields. Fig.~\ref{fig:compare}b shows that $\epsilon_{energy}$ also increases almost monotonically with $a_0$ (purple points). At $a_0 > 40$, the errors in the energy gain become significant, with the maximum energy underestimated by a factor of three at $a_0=100$. 

The results for the perfectly propagating PIC rule out the numerical wave propagation as the primary cause for the poor performance at $a_0 \gg 1$. The observed errors are clearly associated with the particle integrator. It was previously shown that the Boris algorithm itself performs poorly even when the fields acting on the particle are prescribed using the analytical form without any temporal interpolation~[\onlinecite{Arefiev_PoP_2015-resolution}]. In this case, the analytical solution provides \gc{$\bm{E}_{n}$ and $\bm{B}_{n}$} directly for the Boris algorithm. It was found that the errors originate at the stopping points of the trajectory where large rotations of the momentum are performed in a single time-step~[\onlinecite{Arefiev_PoP_2015-resolution}]. It was shown that the errors in the energy gain can be dramatically reduced by sub-cycling the Boris algorithm's time-steps near the stopping points~[\onlinecite{Arefiev_PoP_2015-resolution}]. 

In an attempt to improve the performance of the considered PIC algorithms, we first implement a similar sub-cycling algorithm as described in Ref.~[\onlinecite{Arefiev_PoP_2015-resolution}]. \gc{The details of implementation are shown as Algorithm~\ref{alg1} below.} We define a maximum rotation angle, $\Psi_{max}$, that the particle's momentum is allowed to experience during that single time-step. Conveniently, the Boris algorithm splits the evolution of the particle's momentum into three steps: half of an electric field push, a magnetic field rotation, and another half of an electric field push. We can thus use that rotation generated by the magnetic field to determine the size of the sub-cycling time-step $\Delta t^*$ for the Boris algorithm. The momentum rotation over the time-step $\Delta t^*$ is
\begin{equation}
    \Psi = -\frac{|e| B \Delta t^*}{2 \gamma m c}.
\end{equation}
We set $\Delta t^* = \Delta t/4^k$ where $k$ is the smallest whole number that satisfies the condition $|\Psi| < \Psi_{max}$, with $\Delta t$ being the original time-step and also the time-step for the Yee scheme. 

The switch to a new time-step must be consistent with the leapfrog scheme used to advance the particle's position in the standard PIC algorithm. In order to keep the momentum and position of the sub-cycled particle staggered by half a time-step, the first momentum push when changing $\Delta t^*$ from $\Delta t^*_{old}$ to $\Delta t^*_{new}$ must be performed using $\Delta t^* = (\Delta t^*_{old} + \Delta t^*_{new})/2$ \gc{(see the for-loop in Algorithm~\ref{alg1})}. The position and momentum updates that follow should then use  $\Delta t^* =\Delta t^*_{new}$. This is a general approach, so it equally applies to those cases where $\Delta t^*_{old} = \Delta t$ or $\Delta t^*_{new} = \Delta t$.

\begin{algorithm} 
\SetInd{0.25em}{0.5em}
\DontPrintSemicolon
\caption{Sub-Cycling} \label{alg1}
    $\Delta t^*_{old} = \Delta t^*_{new}$ ($\Delta t^*_{new}$ from previous $\Delta t$ time-step)\;
    $\Delta t^*_{new} = \Delta t$\;
    \While{$|\Psi| > \Psi_{\max}$}{
    \Indp \Indp \Indp $\Delta t^*_{new} = \Delta t^*_{new} / 4$\;
    $\Psi = - |e| B \Delta t^*_{new} / 2 \gamma m c$;}
    $counter=(\Delta t^*_{old} / \Delta t^*_{new})$\;
    \For{$j=1:counter$}{
    \Indp \Indp \Indp $\Delta t^*=(\Delta t^*_{old}+\Delta t^*_{new})/2$\;
    find $\bm{p}_{j+1/2}$ from Eq.~(\ref{eq:p})\;
    \Indp \Indp using $\Delta t^*$ instead of $\Delta t$\;
    \Indm \Indm $\Delta t^*=\Delta t^*_{new}$\;
    find $\bm{x}_{j+1}$ from Eqs.~(\ref{eq:gamma}) - (\ref{eq:x})\;
    \Indp \Indp using $\Delta t^*$ instead of $\Delta t$\;
    \Indm \Indm $\Delta t^*_{old} = \Delta t^*_{new}$\;}
\end{algorithm}


In order to synchronize the motion of sub-cycled particles with the field discretization, we set, as stated earlier, $\Delta t^* = \Delta t/4^k$, where $k$ can be different for different particles. The temporal and spatial step sizes for the fields are fixed to $\Delta t$ and $\Delta x$ for the duration of the entire simulation. The sub-cycling algorithm is used only to advance particles, with the $|\Psi| < \Psi_{max}$ condition applied to each electron individually to determine the corresponding $\Delta t^*$. The condition is only applied once at the beginning of every $\Delta t$ time-step to determine the value of $\Delta t^*$ for the whole $\Delta t$ time-step. Since the fields are not known at the sub-cycled time-steps $\Delta t^*$, we use a linear interpolation procedure for both the electric and magnetic fields. Specifically, we interpolate the electric and magnetic field fields to $t + \Delta t^* / 2$ in order to advance the momentum from $t$ to $t + \Delta t^*$. \gc{This allows the Boris algorithm to update the momentum using field values halfway of each $\Delta t^*$.} \gc{Spatially, the fields are calculated by the particle's triangular shape function}. Once the entire time interval $\Delta t$ has been sub-cycled for each electron, the fields must be updated and individual time-steps $\Delta t^*$ can be changed again to satisfy the new $|\Psi| < \Psi_{max}$ rotation condition. 


\begin{figure}
    \centering
    \includegraphics[width=\columnwidth]{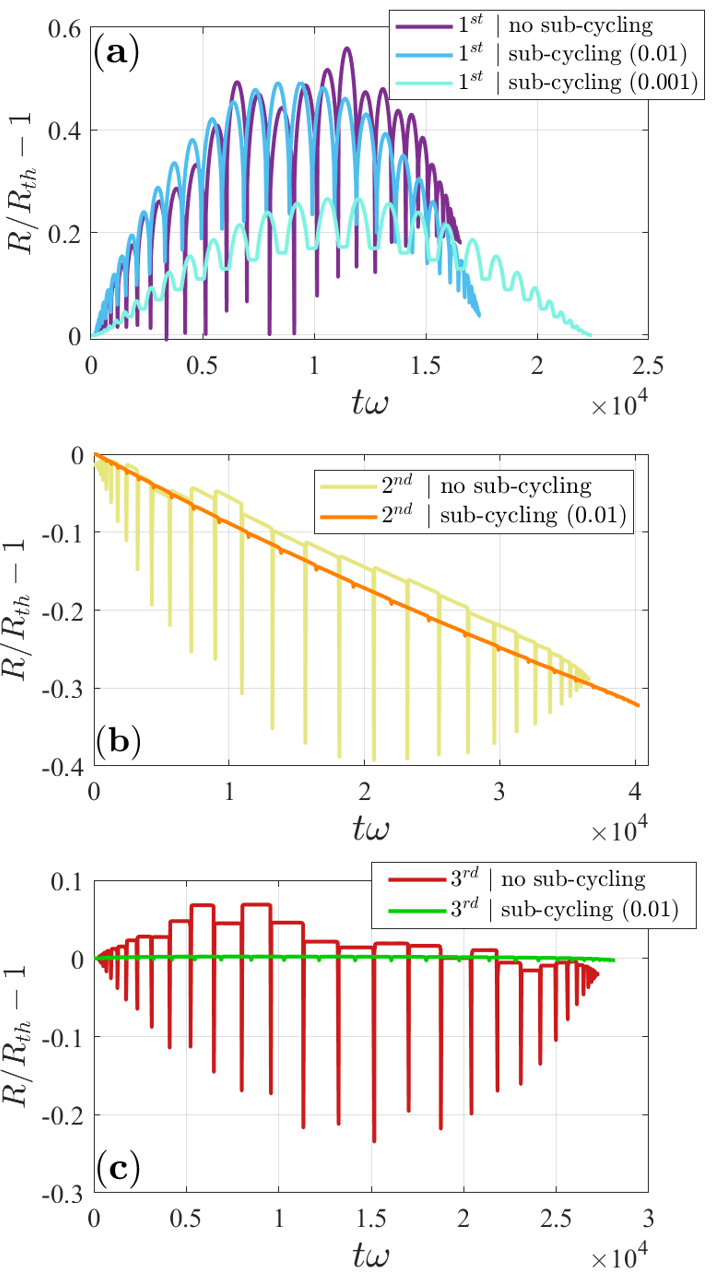}
    \caption{Dephasing error along the electron trajectory in the model problem at $a_0=50$ for different temporal field interpolation orders [$1^{st}$ (a), $2^{nd}$ (b), and $3^{rd}$ (c)] in perfectly propagating PIC with and without the sub-cycling. All simulations use $\Delta t = \lambda / 150 c$ \gc{and $\Delta x \approx \lambda / 148.52$ to satisfy $c \Delta t / \Delta x  = 0.99$}. The maximum rotation angle $\Psi_{max}$ for the sub-cycling is shown in parentheses.}
    \label{fig:traj}
\end{figure}

Fig.~\ref{fig:compare} shows the results from the standard and perfectly propagating PIC algorithm that employ the described sub-cycling with $\Psi_{max}=0.01$ (blue markers). The time-step $\Delta t$ and the grid-size $\Delta x$ are the same as the ones used for the simulations without the sub-cycling (purple markers). It is unclear if the performance is better when using the sub-cycling. Figures~\ref{momentum}a and \ref{momentum}b illustrate the electron trajectory and momentum obtained using the standard PIC with sub-cycling at $a_0 = 100$. The trajectory is symmetric, but it is greatly distorted, whereas the maximum energy gain is roughly three times lower than what it should be according to the analytical solution. The fact that the sub-cycling delivers a dramatic improvement only in the case when the fields are prescribed analytically at the electron location and at the exact time of the momentum update~[\onlinecite{Arefiev_PoP_2015-resolution}] strongly suggests that there are additional sources of error associated with the temporal interpolation. 


\section{Impact of temporal interpolation on the particle integrator} \label{Sec-5} 

To start this section, we look to qualitatively understand why the use of the sub-cycling algorithm delivers only marginal improvement. We examine the time evolution of the dephasing rate during a single simulation of the model problem with the perfectly propagating PIC. In Fig.~\ref{fig:traj}a, the relative dephasing error, $(R - R_{th})/R_{th}$, is shown for every time-step. The purple curve represents a simulation without the sub-cycling, whereas the blue curve represents a simulation with the sub-cycling using $\Psi_{max}=0.01$. The sharp downward spikes along these curves coincide with the stopping points along the electron trajectory. Even though the sub-cycling curve shows noticeable decrease in the length of the spikes, it still exhibits a significant departure from $R_{th} = 1$, which is similar to that of the curve without the sub-cycling. 


In this section, we will show that the just mentioned accumulation of the dephasing error is associated with temporal interpolation of temporally discretized electric and magnetic fields. As explained in Sec.~\ref{Sec-extra}, a common approach is to use linear or first order temporal interpolation for the magnetic field. In the case of sub-cycling, we use the same interpolation procedure for the electric field. For the rest of the paper, this approach will be referred to as the $1^{st}$ order interpolation.

\begin{figure}
    \centering
    \includegraphics[width=\columnwidth]{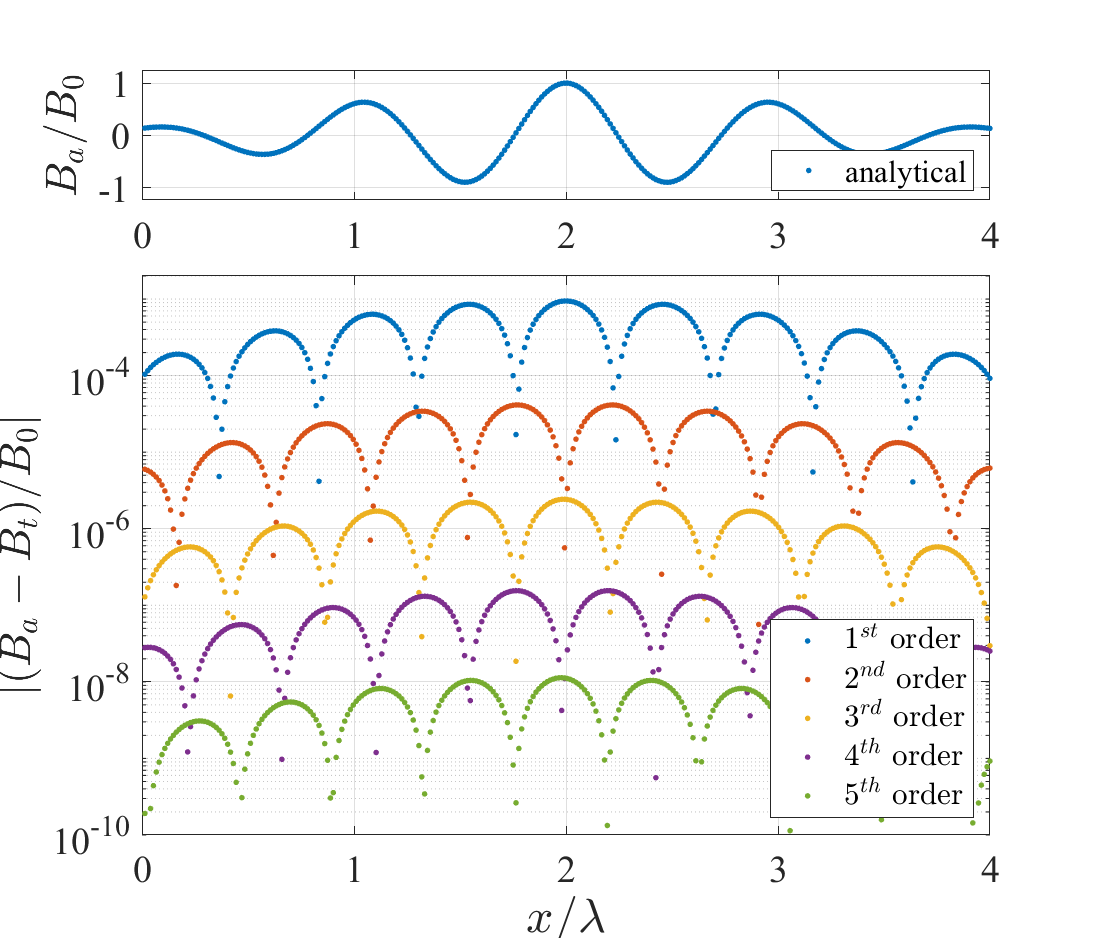}
    \caption{Errors induced by $1^{st}$ to $5^{th}$ order temporal interpolation in the perfectly propagating PIC. Upper panel: magnetic field $B_a$ of the pulse. Lower panel: interpolation error, where $B_t$ is interpolated to $t = n \Delta t$ using $B_a$ at $t = (n + 1/2) \Delta t$ and at preceding time-steps. The temporal discretization is performed using $\Delta t = \lambda / 75 c$. 
    }
    \label{fig:tint}
\end{figure}

An example of the error produced by the $1^{st}$ order interpolation in the perfectly propagating PIC algorithm is shown in the lower panel of Fig.~\ref{fig:tint}. $B_a/B_0$ is shown in the upper panel, where $B_a$ is the analytical solution for the magnetic field of a propagating laser while $B_0>0$ is the peak amplitude. We evaluate the error by plotting $|(B_a - B_t)/B_0|$, where $B_t$ is the temporally interpolated field using the analytical field $B_a$ known at a half time-step before and after on a fixed spatial grid. The biggest errors occur close to the maxima and minima of $B_a$ where the field has a quadratic rather than linear dependence on the phase variable $\xi$ and thus time $t$.

Though the errors shown in Fig.~\ref{fig:tint} are seemingly small, they can have a profound impact on the energy gain by altering the electron dynamics. We illustrate this aspect by considering the temporal field discretization used by the perfectly propagating PIC without the sub-cycling. This algorithm uses an analytically propagated field that requires a $1^{st}$ order temporal interpolation of the magnetic field to advance the electron momentum. Analytically, as given by Eq.~(\ref{laser5}), the magnetic field acting on the electron is equal to the electric field. This is because the laser propagation is in a vacuum, which results in the phase velocity, $v_{ph}$, being equal to the speed of light, i.e. $v_{ph} = c$. Problematically, the $1^{st}$ order temporal interpolation effectively reduces the magnetic field amplitude. As a result, this interpolation makes the field configuration similar to that of a laser pulse with a superluminal phase velocity:
\begin{equation} \label{superlum}
    B = E c / v_{ph}.
\end{equation}
It can be shown that the integral of motion in such a pulse~[\onlinecite{Univ.ScalingIonChannel2016}] is
\begin{equation} \label{superlum constant}
    \gamma - \frac{v_{ph}}{c} \frac{p_x}{m c} = \mbox{const}
\end{equation}
rather than $\gamma - p_x/mc = \mbox{const}$. Eq.~(\ref{superlum constant}) is a manifestation of the changes in the electron dynamics due to the interpolation. 

We now use the analogy between the interpolated field and that of a superluminal laser pulse to estimate the induced error in the dephasing $R$. For simplicity, we neglect spatial discretization in the derivation that follows. The extrapolated field is
\begin{equation}
    B_t = B_a - \Delta B,
\end{equation}
so that, after taking into account that $E = B_a$, we obtain
\begin{equation}
    B_t / E = 1 - \Delta B / B_a.
\end{equation}
It then follows from Eq.~(\ref{superlum}) that the expression on the right-hand side can be interpreted as $c/v_{ph}$, such that the effective phase velocity is given by
\begin{equation} \label{Eq:v_ph}
    v_{ph}/c \approx 1 + \Delta B / B_a.
\end{equation}
Note that the second term on the right-hand side is never negative during the $1^{st}$ order interpolation, so that $v_{ph}/c \geq 1$. We substitute the expression given by Eq.~(\ref{Eq:v_ph}) into Eq.~(\ref{superlum constant}) to find that
\begin{equation} \label{superlum constant 2}
    \gamma - \frac{p_x}{m c} = \mbox{const} + \frac{\Delta B}{B_a} \frac{p_x}{m c}. 
\end{equation}
Recall that we are using the analytical form for the fields, so the dephasing defined by Eq.~(\ref{R-def}) is equal to $R = \gamma - p_x/m c$. It then follows from Eq.~(\ref{superlum constant 2}) that the reduction in the magnetic field strength due to the interpolation, $\Delta B/B_a > 0$, increases the dephasing between the electron and the wave,
\begin{equation} \label{interp dephasing}
    R = \gamma - \frac{p_x}{m c} = \mbox{const} + \frac{\Delta B}{B_a} \frac{p_x}{m c}.
\end{equation} 
An important consequence of this result is that even small interpolation errors can lead to significant errors in the dephasing for an electron with an ultra-relativistic longitudinal momentum, i.e. $p_x \gg mc$.

\begin{figure}
    \centering
    \includegraphics[width=\columnwidth]{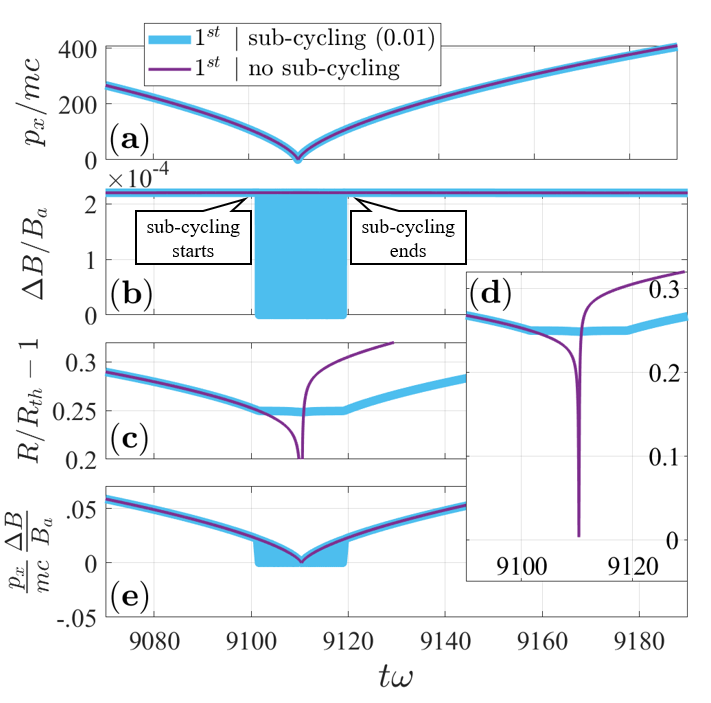}
    \caption{Performance of perfectly propagating PIC using the $1^{st}$ order interpolation with and without the sub-cycling near a stopping point. (a): longitudinal electron momentum. (b): the error between the interpolated magnetic field and the analytical magnetic field. (c) and (d): relative error in the dephasing rate. (e): prediction for the relative error in the dephasing rate given by Eq.~(\ref{interp dephasing}). The sub-cycling ($\Psi_{max}=0.01$) is only performed near the stopping point shown in these plots. Both simulations use $a_0=50$, $\Delta t = \lambda / 150 c $ \gc{and $\Delta x \approx \lambda / 148.52$.}}
    \label{fig:local}
\end{figure}

In order to validate the mechanism linking the temporal interpolation with the dephasing errors, we re-examine the plot from Fig.~\ref{fig:traj}a for the algorithm without the sub-cycling. Fig.~\ref{fig:local} zooms in around a stopping point at $t \omega \approx 9110$. Fig.~\ref{fig:local}c shows the relative dephasing error $(R - R_{th})/R_{th}$, whereas Fig.~\ref{fig:local}e shows the error that we expect due to the temporal interpolation according to Eq.~(\ref{interp dephasing}). In order to be consistent with the general PIC approach, we apply the triangular particle shape to calculate both $B_a$ and $\Delta B$. At $t \omega < 9100$, Eq.~(\ref{interp dephasing}) reproduces the temporal dependence of $(R - R_{th})/R_{th}$ remarkably well, which confirms our estimates and the assertion that most of the errors at $p_x/mc \gg 1$ are caused by the interpolation.  

Figures~\ref{fig:local}c and \ref{fig:local}d show how the sub-cycling removes the sharp downward spike and the noticeable asymmetry in $(R - R_{th})/R_{th}$ at the stopping point. To aid the comparison, we only use the sub-cycling for the considered stopping point, i.e. no sub-cycling is applied at $t \omega < 9100$. The curve becomes noticeably flatter during the sub-cycling, which suggests that the sub-cycling procedure mitigates the negative impact of the $1^{st}$ order temporal interpolation. 

It must be pointed out that the sub-cycling algorithm inherits the field interpolation error. Temporal interpolation during the sub-cycling causes $\Delta B / B_a$ to rapidly oscillate (see Fig.~\ref{fig:local}b). To understand the nature of the oscillations, recall that the magnetic field is known at $n - 1/2$ and $n + 1/2$, which means that there is no magnetic field interpolation error at those time indices. The result is that $\Delta B$ during the sub-cycling oscillates between 0 and some interpolation errors.

We further investigate the impact of the sub-cycling by applying the sub-cycling algorithm with $\Psi_{max}=0.001$ to the entire simulation. The result is shown in Fig.~\ref{fig:traj}a. We do see a noticeable improvement compared to the simulation with $\Psi_{max}=0.01$. However, the number of time integration steps increases by a factor of 4 to roughly $2.1 \times 10^6$ steps. In terms of the number of sub-cycling steps, the increase is by a factor of 21 from 72,000 steps to $1.5 \times 10^6$ steps. Even though the sub-cycling does solve some local problems at the stopping points, it is unable to prevent accumulation of $(R - R_{th})/R_{th}$ even when using a computationally expensive $\Psi_{max}$ in the sub-cycling algorithm. This accumulation leads to significant deviations from the analytical solution (for example, see Figs.~\ref{momentum}a and \ref{momentum}b for $\Psi_{max}=0.01$). 

We then conclude that, even with the use of sub-cycling, the $1^{st}$ order temporal field interpolation has the effect that is similar to that of a superluminal laser pulse, resulting siginificant errors for ultra-relativistic particles. The effective superluminosity is particularly evident from the momentum plot in Fig.~\ref{momentum} that resembles that for an electron in a plane wave with $v_{ph}>c$~[\onlinecite{Robinson_PoP_2015}].


\section{Higher-order field interpolation} \label{Sec-6} 

In order to reduce the errors caused by the $1^{st}$ order (linear) temporal interpolation of the fields, we change the interpolation procedure by increasing the order of the polynomial using Lagrange interpolation. As discussed later in this section, we find that a $2^{nd}$ order interpolation produces only marginal improvements whereas a $3^{rd}$ order interpolation dramatically reduces numerical errors. 

The $1^{st}$ order interpolation for the magnetic field without the sub-cycling has the form
\begin{equation}
    B_z|_{n} = \frac{1}{2} \left( B_z |_{n + 1/2} + B_z |_{n - 1/2} \right).
\end{equation}
However, the information about the magnetic field is known at $t < (n-1/2) \Delta t$, so we can leverage it to more accurately describe $B_{z}|_{n}$:
\begin{eqnarray}
    &&B_z |_n = \sum_{j=0}^{p} B_z |_{n + 1/2 + j - p} \mathcal{L}_{p,j}, \label{eq:lagrange1}
\end{eqnarray}
where $p$ is the order of the interpolation and $\mathcal{L}_{p,j}$ are the Lagrange coefficients defined as
\begin{eqnarray}
    &&\mathcal{L}_{p,j}=\prod_{k=0,k \neq j}^{p} \frac{p-k-1/2}{j-k}. \label{eq:lagrange2}
\end{eqnarray}
For example, the $2^{nd}$ order interpolation is given by
\begin{eqnarray}
    &&B_z |_n = \frac{1}{8} \left( -B_z|_{n - 3/2} + 6 B_z|_{n - 1/2} + 3 B_z|_{n + 1/2} \right). \label{eq:2nd interp}
\end{eqnarray}

In order to apply the interpolation during the sub-cycling, we generalize Eqs.~(\ref{eq:lagrange1}) and (\ref{eq:lagrange2}) to the case when the field is interpolated to $t = \Delta t (n-1/2) + \Delta t'$. Then the corresponding field that we denote as $B'_z$ is given by
\begin{eqnarray}
    &&B'_z = \sum_{j=0}^{p} B_z|_{n + 1/2 + j - p} \mathcal{L}^B_{p,j}, \label{eq:lagrange1-2}
\end{eqnarray}
where
\begin{eqnarray}
    &&\mathcal{L}^B_{p,j}=\prod_{k=0,k \neq j}^{p} \frac{p-k-1 + \Delta t' / \Delta t}{j-k}. \label{eq:lagrange2-2}
\end{eqnarray}
These expressions are used for $\Delta t^* \leq \Delta t' \leq \Delta t$, where $\Delta t^*$ is the sub-cycling step. As previously stated, the sub-cycling algorithm also requires the electric field to be temporally interpolated. The interpolated expression at $t = \Delta t (n-1/2) + \Delta t'$ is given by
\begin{eqnarray}
    &&E'_y = \sum_{j=0}^{p} E_y|_{n + j - p} \mathcal{L}^E_{p,j}, \label{eq:lagrange1-3}
\end{eqnarray}
where
\begin{eqnarray}
    &&\mathcal{L}^E_{p,j}=\prod_{k=0,k \neq j}^{p} \frac{p-k-1/2 + \Delta t' / \Delta t}{j-k}. \label{eq:lagrange2-3}
\end{eqnarray}

We first implement the $2^{nd}$ order temporal field interpolation to the perfectly propagating PIC algorithm as described by Eq.~(\ref{eq:2nd interp}). The dephasing errors for simulations with and without the sub-cycling are shown in Fig.~\ref{fig:traj}b. The simulation setup and parameters are the same as those used in Fig.~\ref{fig:traj}a, which uses the $1^{st}$ order interpolation instead. The use of the $2^{nd}$ order interpolation together with the sub-cycling significantly reduces the length of the downward spikes at the stopping points. However, there is still a large error accumulation over the duration of the entire simulation. This behavior is qualitatively different from what we observe for the $1^{st}$ order interpolation where $R/R_{th} - 1$ returns towards zero as the laser pulse overtakes the electron. The error accumulation is caused by the asymmetry of the $2^{nd}$ order interpolation that uses two data points from the ``past'' ($n-3/2$ and $n-1/2$) and only one data point from the ``future'' ($n+1/2$). To test this, we changed the interpolation procedure by using two data points from the ``future'' ($n+1/2$ and $n+3/2$) and only one data point from the ``past'' ($n-1/2$). The resulting $R/R_{th} - 1$ curve also exhibits significant error accumulation, but now $R/R_{th} - 1$ increases rather than decreases.

Our results show that an even higher order interpolation is necessary to improve the accuracy of the simulations. We find that the $3^{rd}$ order temporal interpolation of the fields is sufficient to achieve a dramatic improvement. As shown in Fig.~\ref{fig:traj}c, the dephasing error is reduced dramatically when the $3^{rd}$ order interpolation is used together with the sub-cycling. Increasing the order of the interpolation does not resolve the asymmetry issue, but it does reduce the magnitude of the errors (see lower panel of Fig.~\ref{fig:tint}). 

The role of the higher-order temporal interpolation is to reduce the temporal interpolation errors. The curve for the simulation without the sub-cycling in Fig.~\ref{fig:traj}c shows that the $3^{rd}$ order interpolation makes the error accumulations mostly happen near the stopping points. This means that the interpolation errors are no longer the main source of the dephasing error. This conclusion is reaffirmed as we see that the errors near the stopping points are mitigated when adding the sub-cycling algorithm to the $3^{rd}$ order interpolation. The resulting $R$ in Fig.~\ref{fig:traj}c is very close to $R_{th}$ along the entire trajectory, with very slight downward ticks at the stopping points. Figures~\ref{momentum}a and \ref{momentum}b confirm that the PIC algorithm that uses both the sub-cycling and the $3^{rd}$ order interpolation is able to recover the electron trajectory and the momentum evolution.

We conclude this section by performing a broad laser amplitude scan by increasing $a_0$ from 5 to 100. In Fig.~\ref{fig:compare}b, the green markers show $\epsilon_{energy}$ for a perfectly propagating PIC that uses both the sub-cycling and the $3^{rd}$ order interpolation. The errors are dramatically reduced compared to the algorithm that uses the $3^{rd}$ order interpolation without the sub-cycling. Fig.~\ref{fig:compare}a confirms that a similar trend is in place for the standard PIC algorithm where the fields are numerically propagated. These scans show that a higher order interpolation is necessary for the sub-cycling algorithm to yield significant benefits.


\section{Improvements due to $3^{rd}$ and $5^{th}$ order temporal interpolation} \label{Sec-7}

\begin{figure}
    \centering
    \includegraphics[width=\columnwidth]{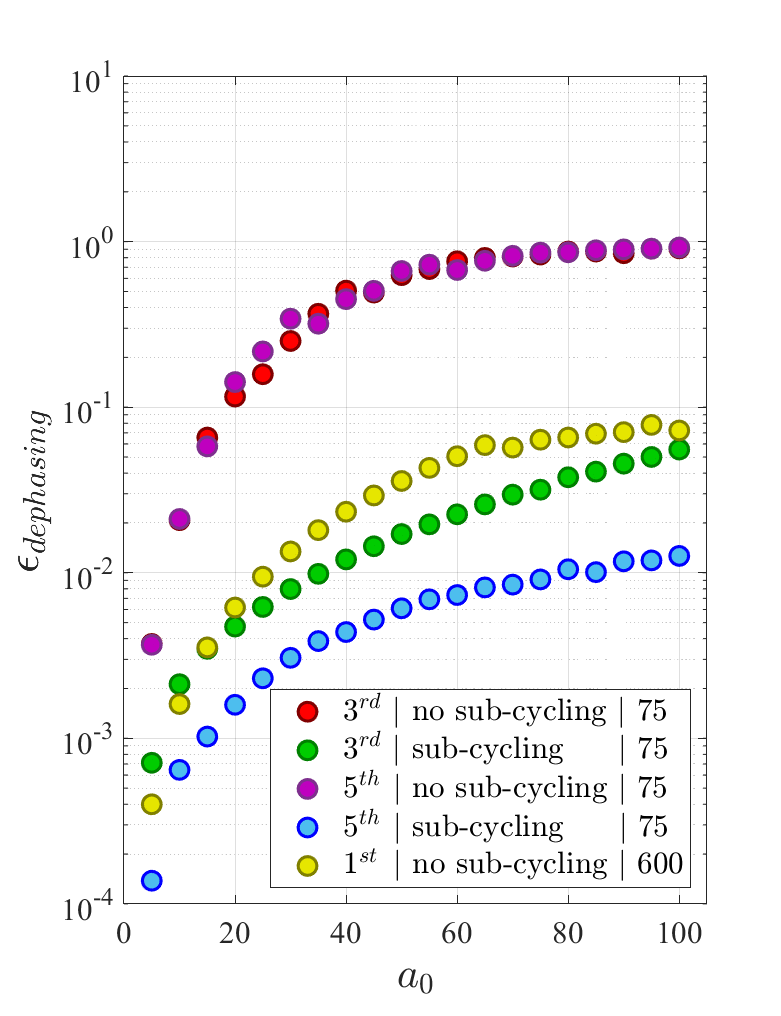}
    \caption{Relative dephasing error, $\epsilon_{dephasing}$, for different versions of the PIC algorithm over a range of laser amplitudes $a_0$. All simulation results are obtained using the standard PIC algorithm. Each data set represents simulations using a different combination of temporal field interpolation orders, with/without sub-cycling, and discretization resolution. All simulations use \gc{$c \Delta t / \Delta x  = 0.99$} and $\Psi_{max}=0.01$ is with sub-cycling.}
    \label{compare_more}
\end{figure}

In this section, we quantify improvements in precision delivered by implementing a higher-order temporal field interpolation and the sub-cycling into a standard PIC algorithm. We compare the results for $3^{rd}$ and $5^{th}$ order interpolations to those obtained using the standard $1^{st}$ order interpolation.

Fig.~\ref{compare_more} shows $\epsilon_{dephasing}$ as a function of $a_0$ for simulations using the $5^{th}$ order temporal interpolation with and without the sub-cycling. The $3^{rd}$ order points in this figure are exactly the same points as found in Fig.~\ref{fig:compare}a. Without the sub-cycling, there is essentially no change in $\epsilon_{dephasing}$ when increasing the interpolation order from $3^{rd}$ to $5^{th}$. When the algorithm employs the sub-cycling, the $5^{th}$ order interpolation does reduce the errors compared to the $3^{rd}$ order interpolation. 

\rc{Continuing in Fig.~\ref{compare_more},} it is instructive to determine the resolution needed in the \gc{standard PIC algorithm} ($1^{st}$ order interpolation and no sub-cycling) to achieve \rc{low values} of $\epsilon_{dephasing}$. \rc{We use the results of the new algorithm ($3^{rd}$ order interpolation with sub-cycling, shown by the green points) for comparison. Scanning over $5 \leq a_0 \leq 100$ and using a resolution of $\Delta t = \lambda / 75 c$, the new algorithm maintained $\epsilon_{dephasing} < 0.1$ throughout.} \rc{We found that, in the standard PIC algorithm, $\Delta t = \lambda / 600 c$ is required to achieve similar results (shown by the yellow points)}. We kept \gc{$c \Delta t / \Delta x  = 0.99$} during \rc{both of these} scans. Therefore, not only the number of time-steps has increased by a factor of 8, but the number of cells has also increased by the same factor. This comparison shows that improving the precision of the \gc{standard PIC algorithm} through a straightforward reduction of $\Delta t$ and $\Delta x$ can become prohibitively expensive. 

\begin{figure}[t]
    \centering
    \includegraphics[width=\columnwidth]{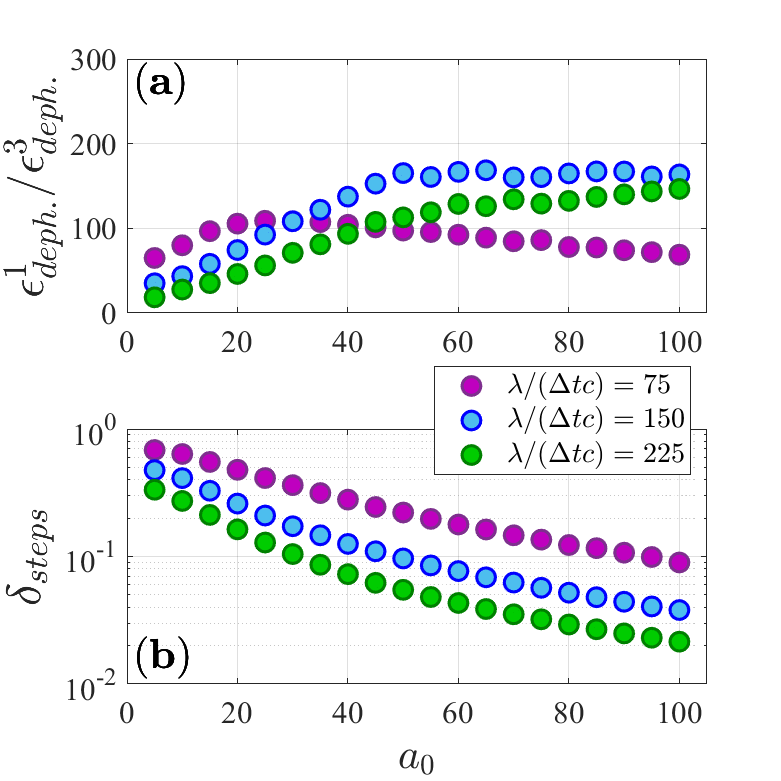}
    \caption{Improvements due to the $3^{rd}$ order interpolation, $\epsilon_{dephasing}^{1} / \epsilon_{dephasing}^{3}$, and required increase in the number of time steps, $\delta_{steps}$, for PIC algorithm with sub-cycling ($\Psi_{max}=0.01$). All simulations use $c \Delta t / \Delta x  = 0.99$.  }
    \label{fig:error3rd}
\end{figure}


\begin{figure}[t]
    \centering
    \includegraphics[width=\columnwidth]{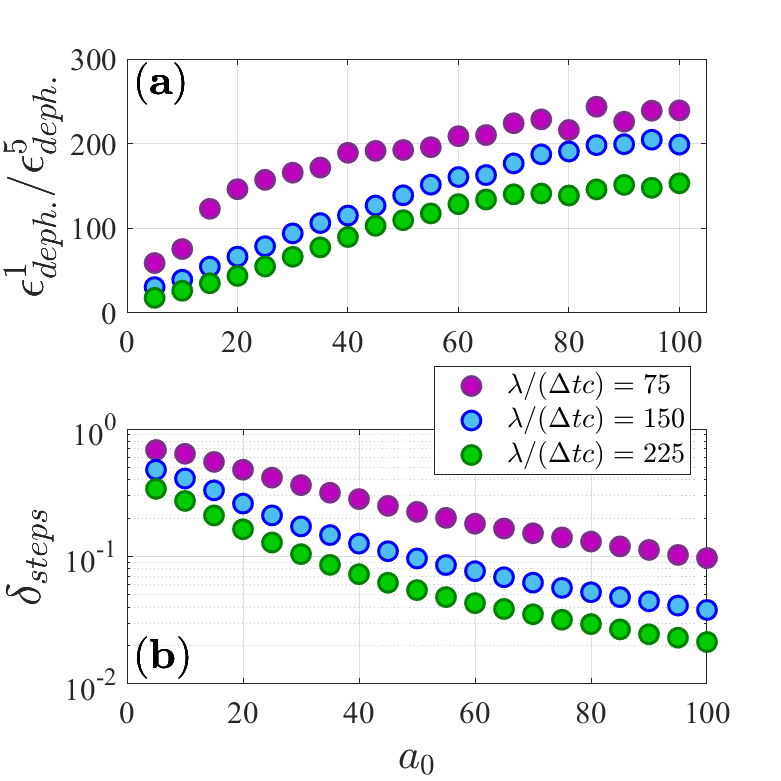}
    \caption{Improvements due to the $5^{th}$ order interpolation, $\epsilon_{dephasing}^{1} / \epsilon_{dephasing}^{5}$, and required increase in the number of time steps, $\delta_{steps}$, for PIC algorithm with sub-cycling ($\Psi_{max}=0.01$). All simulations use \gc{$c \Delta t / \Delta x  = 0.99$}. }
    \label{fig:error5th}
\end{figure}

To have a more direct comparison for how important the high order temporal field interpolation is in ensuring effective sub-cycling, we introduce
\begin{eqnarray}
    && \chi^{3} = \left. \epsilon_{dephasing}^{1} \right/ \epsilon_{dephasing}^{3} \\
    && \chi^{5} = \left. \epsilon_{dephasing}^{1} \right/ \epsilon_{dephasing}^{5},
\end{eqnarray} 
where the superscripts indicate the order of the temporal interpolation in a simulation with the sub-cycling. Figures~\ref{fig:error3rd} and \ref{fig:error5th} show $\chi^{3}$ and $\chi^{5}$ as functions of $a_0$ for three different values of $\lambda/(\Delta t c)$ which denotes the resolution of both $\epsilon_{dephasing}^{1}$ and $\epsilon_{dephasing}^{3}$ in the $\chi^{3}$ calculation. 

A significant feature of Fig.~\ref{fig:error3rd}a is the maximum of $\chi^{3}$ at $a_0 \approx 25$ for $\Delta t = \lambda / 75 c$. The location of the maximum shifts to higher values of $a_0$ as we increase the resolution. The implication of this trend is that the resolution must be adjusted according to the value of $a_0$ in order to maintain the effectiveness of the higher order interpolation.

We find that $\chi^{3}$ is reduced with the increase of $a_0$ primarily during the ultra-relativistic electron motion. This was confirmed by first observing that an almost identical trend exists for the perfectly propagating PIC. We then replaced the fields in the perfectly propagating PIC by their analytical values without the interpolation along the ultra-relativistic parts of the trajectory, which caused $\chi^{3}$ to become monotonically increasing. We note that $c - v_x$ decreases with the increase of $a_0$, where $v_x$ is the longitudinal electron velocity along the ultra-relativistic part of the trajectory. The electron motion becomes more sensitive to errors in the effective phase velocity caused interpolation as we increase $a_0$ (see Eq.~\ref{interp dephasing}).

An alternative to increasing the resolution is to increase the order of the interpolation. Fig.~\ref{fig:error5th}a shows that the interpolation errors are reduced sufficiently to prevent a rollover of $\chi^{5}$ at $a_0 \leq 100$ for $\Delta t = \lambda / 75 c$. The penalty for using the $5^{th}$ order interpolation is the need to store more information about the electric and magnetic fields. However, one can use relatively large cells and time-steps, the saving of which might be significant when using this approach for 3D simulations. 

When the resolution $\lambda/(\Delta t c)$ becomes more fine in the improved PIC algorithm, $\epsilon_{dephasing}$ unsurprisingly decreases. Though we see in Fig.~\ref{fig:error5th}a that $\chi^5$ is reduced with larger $\lambda/(\Delta t c)$. This shows that simulations with coarser resolutions have the most potential to be improved by increasing the order of the temporal field interpolation given that the interpolation is sufficiently high-order. 

Figures~\ref{fig:error3rd} and \ref{fig:error5th} also show the relative increase in the number of time-steps when using the sub-cycling. We define
\begin{equation}
    \delta_{steps}=\frac{N^*}{N^*+N},
\end{equation}
where  $N$ is the number of $\Delta t$ sized time-steps taken during the simulation. $\Delta t$ is the fixed time-step size of the Yee scheme and the without sub-cycling Boris algorithm. $N^*$ is the number of sub-cycling time-steps $(\Delta t^* < \Delta t)$ taken during the simulation. As expected,  by using the same sub-cycling condition, $\delta_{steps}$ is mostly unaffected because the interpolation error $\chi$ is small in the considered examples and the electron trajectories are very similar for the $3^{rd}$ and $5^{th}$ order interpolations. 

As $a_0$ increases, $\delta_{steps}$ decreases. This happens because the travel time between the stopping points increases with $a_0$. The sub-cycling is applied only in the vicinity of the stopping points, which means that the sub-cycling has to be applied less frequently at higher $a_0$. At finer resolutions, $\delta_{steps}$ also decreases as $\Psi$ is scales proportionally with the time-step size.

\section{Summary and Discussion} \label{discuss}

In the 1D PIC tests where the errors induced by the field solver are negligible, the standard linear interpolation of the fields \gc{temporally} can introduce significant errors to the motion of charged particles. These errors can be interpreted as an effective increase in the phase velocity of the laser fields, and are strongly pronounced in the ultra-relativistic regime. We have shown that a high order temporal interpolation for electric and magnetic fields is required in addition to particle sub-cycling near stopping points to accurately reproduce electron dynamics in ultra-high intensity laser pulses. \rc{The need for the high-order interpolation is a general requirement for simulations at $a_0 \gg 1$, so our findings equally apply to implementations that employ a different particle pusher with an improved performance at $a_0 \gg 1$, such as the one detailed in Ref.~[\onlinecite{GORDON2021_SpecialUnity}].}

We have found that a third rather than second order temporal interpolation is needed in order to fully leverage the benefits of the sub-cycling algorithm and thus dramatically improve the simulation accuracy. While the higher order temporal field interpolations would require increases in memory usage, the sub-cycling allows for fairly coarse discretizations to achieve results comparable to standard simulations using very fine discretizations. 

\gc{In the considered problem, the accuracy of the solution is primarily impacted by the temporal field interpolation and the errors introduced by the particle pusher, while the errors introduced by the field solver are relatively inconsequential. If, in a more general case, the errors from the field solver become significant, then it may be justifiable to consider using a higher order solver. Since the errors introduced by the field solver are separate from the time-interpolation errors, the two issues should be treated separately. }

\section *{Acknowledgements} \label{acknowledge}

The work of K.T. and A.A. was supported by the National Science Foundation (PHY 1821944). G. C.’s work was supported by the Exascale Computing Project (grant no. 17-SC-20-SC), a collaborative effort of the U.S. Department of Energy Office of Science and the National Nuclear Security Administration.

\bibliographystyle{apsrev4-1}
\bibliography{Collection}

\section*{Appendix}

\subsection{Electron Dynamics in a Plane Wave}

The equations of motion for an electron have the form:
\begin{eqnarray}
&& \frac{d \bm{p}}{d t} = - |e| \bm{E} - \frac{|e|}{\gamma m c} \left[ \bm{p} \times \bm{B} \right], \label{EQ_1} \\
&& \frac{d \bm{r}}{d t} = \frac{c}{\gamma} \frac{\bm{p}}{m c}. \label{EQ_2}
\end{eqnarray}
We are considering a plane electromagnetic wave propagating along the $x$-axis in the positive direction. The fields can then be described using a normalized vector potential $a$ that is only a function of a phase variable $\xi$, with
\begin{eqnarray}
    && E_y = B_z = - \frac{m \omega c} {|e|} \thinspace \frac{d a}{d \xi}, \label{AA-fields}\\
    && \xi \equiv \omega (t - x / c).
\end{eqnarray}

The three components of Eq.~(\ref{EQ_1}) now read
\begin{eqnarray}
&& \frac{d p_x}{d t} = - \frac{|e|}{\gamma m c} p_y B_z, \label{AA-1}\\
&& \frac{d p_y}{d t} = - |e| E_y + \frac{|e|}{\gamma m c} p_x B_z, \label{AA-2}\\
&& \frac{d p_z}{d t} = 0. \label{AA-3}
\end{eqnarray}
It follows from Eq.~(\ref{AA-3}) that $p_z$ is conserved. In this work we consider an electron that is initially at rest, so that 
\begin{equation} \label{AA-pz}
    p_z = 0
\end{equation}
during the electron motion in the laser pulse. Next, we substitute the expressions for the fields given by Eq.~(\ref{AA-fields}) into Eq.~(\ref{AA-2}) and take into account that $\bm{v} = \bm{p} / \gamma m$ to find that
\begin{equation} \label{AA-4}
    \frac{d p_y}{d t} = m \omega c \left(  1 - \frac{v_x}{c} \right)  \thinspace \frac{d a}{d \xi}.
\end{equation}
This equation can be simplified even further, because
\begin{equation}
    \frac{d \xi}{dt} = \frac{\partial \xi}{\partial t} + \frac{dx}{dt} \frac{\partial \xi}{\partial x} = \omega \left( 1 - \frac{v_x}{c} \right).
\end{equation}
We use this relation in Eq.~(\ref{AA-4}) to obtain an equation that determines the evolution of $p_y$ for a given vector potential $a$:
\begin{equation}
    \frac{d}{d t} \left( \frac{p_y}{mc} - a \right) = 0.
\end{equation}
We then have
\begin{equation} \label{AA-py}
    p_y/mc = a(\xi)
\end{equation}
for an electron that is immobile prior to the arrival of the laser pulse.

Equations~(\ref{AA-1}) - (\ref{AA-3}) have another integral of motion,
\begin{equation} \label{AA-integral}
    \frac{d}{d t} \left( \gamma - \frac{p_x}{mc} \right) = 0,
\end{equation}
where $\gamma = \sqrt{1 + p^2 / m^2 c^2}$. In order to show this, we add Eq.~(\ref{AA-1}) multiplied by $p_x$ to Eq.~(\ref{AA-2}) multiplied by $p_y$, which yields
\begin{equation}
    \frac{1}{2} \frac{d p^2}{d t} = - |e| E_y p_y .
\end{equation}
Taking into account the definition for $\gamma$, we find that
\begin{equation}
    \frac{d \gamma}{d t} = - \frac{|e| E_y p_y}{\gamma m^2 c^2}.
\end{equation}
On the other hand, it follows from Eq.~(\ref{AA-1}) that
\begin{equation}
    \frac{d p_x}{d t} = - \frac{|e|}{\gamma m c} p_y E_y,
\end{equation}
where we used Eq.~(\ref{AA-fields}) to replace $B_z$ with $E_y$. It can now be directly verified that Eq.~(\ref{AA-integral}) holds. 

Using the same initial conditions as before (immobile electron), we find from Eq.~(\ref{AA-integral}) that
\begin{equation} \label{AA-integral-2}
    \gamma - \frac{p_x}{mc} = 1.
\end{equation}
This relation reduces to an equation for $p_x$ after we take into account that $p_y = amc$ and $p_z = 0$:
\begin{equation}
    \gamma = \left( 1 + \frac{p_x^2}{m^2 c^2} + a^2 \right)^{1/2}.
\end{equation}
The solution of the resulting equation is
\begin{equation} \label{AA-px}
    p_x / mc = a^2 /2.
\end{equation}
Equations~(\ref{AA-px}), (\ref{AA-py}), and (\ref{AA-pz}) fully describe the evolution of the momentum for a given normalized vector potential $a = a(\xi)$. It also follows from Eq.~(\ref{AA-integral-2}) that
\begin{equation}
    \gamma = 1 + a^2 / 2.
\end{equation}


\subsection{Energy Conservation Analysis}

It directly follows from the equations of motion for an electron [see Eq.~(\ref{EQ_1})] that
\begin{equation} \label{EQ_35A}
    \frac{d \varepsilon_k}{dt} = - |e| (
    \bm{E} \cdot \bm{v}),
\end{equation}
where 
\begin{equation}
    \varepsilon_k \equiv (\gamma - 1) m c^2
\end{equation}
is the kinetic energy, $\bm{v} = \bm{p} / \gamma m$ is the electron velocity, and $\gamma = (1 + \bm{p}^2 / m_e^2 c^2)^{1/2}$ is the relativistic factor. In our test problem, the electron is initially at rest, with $\varepsilon_k = 0$ at $t=0$. Then the kinetic energy at time $t$ is given by
\begin{equation} \label{eq_energy_conservation}
    \varepsilon_k (t)  = - |e| \int_0^t (
    \bm{E} \cdot \bm{v}) dt'.
\end{equation}

The particle pusher that we use in this manuscript does not automatically guarantee that Eq.~(\ref{eq_energy_conservation}) is satisfied. It is thus insightful to examine the discrepancy between the work that is done by the laser electric field (right-hand side) and the change in the kinetic energy (left-hand side) in our simulations. In order to aid the comparison, we define the kinetic energy in our algorithm as
\begin{equation}
    \varepsilon^{n+1/2}_k = m c^2 (\gamma^{n+1/2} - 1). \label{eq:ke_nosub}
\end{equation}
The work, $w$, done on the electron by the electric field as the velocity changes from $\bm{v}^{n-1/2}$ to $\bm{v}^{n+1/2}$ is 
\begin{equation}
    w^{n+1/2} =  \frac{q \Delta t}{2} \bm{E}^{n} \cdot (\bm{v}^{n-1/2} + \bm{v}^{n+1/2}) . \label{eq:work_nosub}
\end{equation}
The net work done from the beginning of the simulations is then
\begin{equation}
    \varepsilon^{n+1/2}_{w} \equiv \sum^{n}_{i=0} w^{i+1/2}.
\end{equation}




\begin{figure}[htb]
    \centering
    \includegraphics[width=\columnwidth]{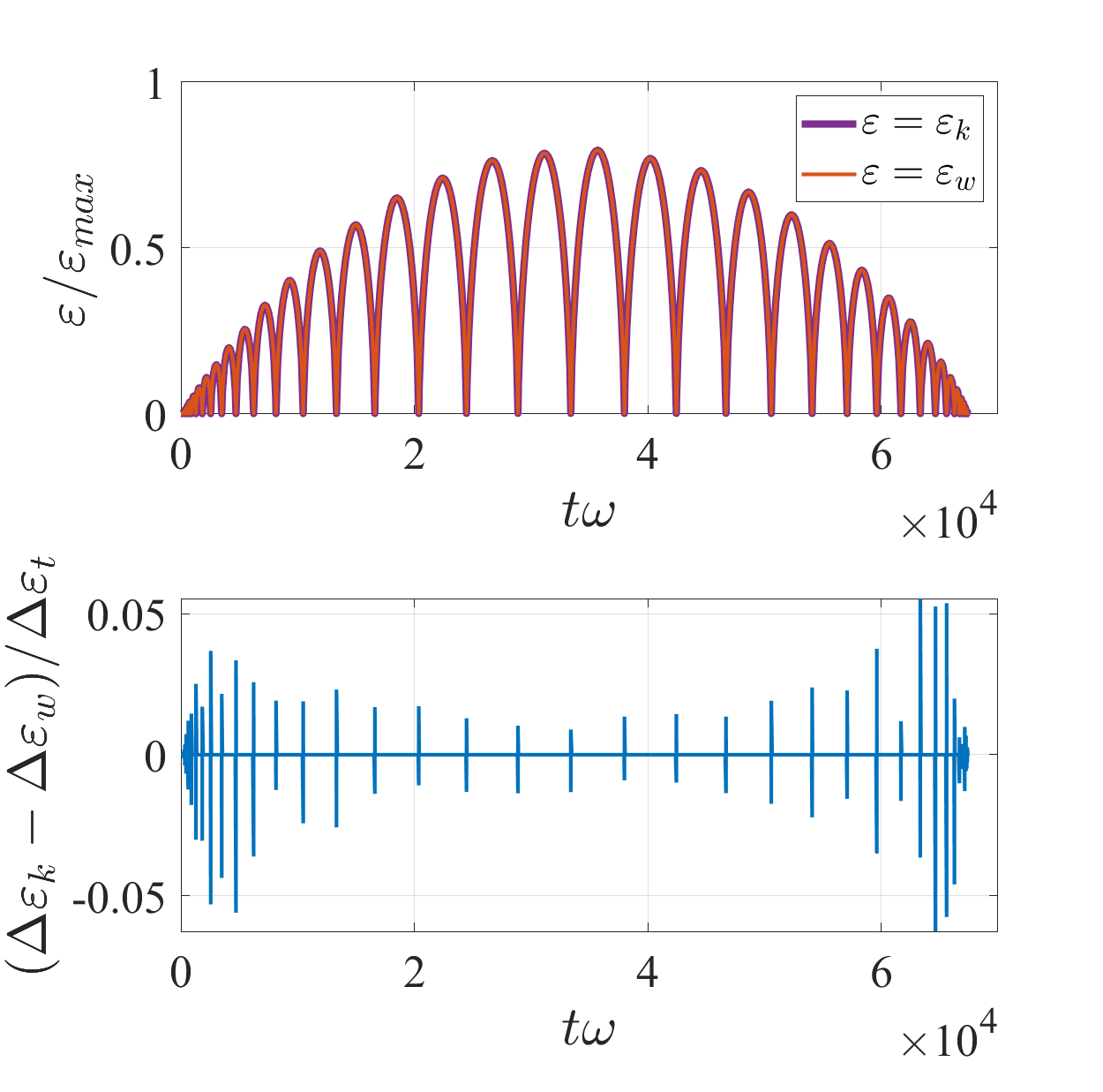}
    \caption{Energy conservation in a PIC algorithm with $3^{rd}$ order temporal interpolation without the sub-cycling algorithm. (top panel) The kinetic energy, $\varepsilon_{k}$, and the net work done, $\varepsilon_{w}$, normalized by $\varepsilon_{max}$. (bottom panel) The difference between the work done and the change in kinetic energy over one $\Delta t$ time-step, $\Delta \varepsilon_{k} - \Delta \varepsilon_{w}$, normalized by $\Delta \varepsilon_{t}$. The simulation parameters are set to $a_0=100$, $\lambda/(\Delta t c) = 75$, and $c \Delta t / \Delta x  = 0.99$. }
    \label{fig:energyconservation_nosub}
\end{figure}

In Fig.~\ref{fig:energyconservation_nosub}, the top panel shows $\varepsilon_k$ and $\varepsilon_w$ in a PIC algorithm that employs the $3^{rd}$ order temporal field interpolation without the sub-cycling. The simulation parameters are provided in the caption. The curves are normalized by $\varepsilon_{max} = m c^2 a_0^2/2$, which is the theoretical maximum kinetic energy of an electron with the considered initial conditions (we neglected the $p_y$ contribution as $p_x\gg p_y$ for $a_0\gg0$). Both $\varepsilon_{k}/\varepsilon_{max}$ and $\varepsilon_{w}/\varepsilon_{max}$ remain significantly below unity along the electron trajectory, which indicates that the algorithm fails to correctly reproduce the energy gain. On the other hand, there is no visible difference between the two curves. We thus conclude that the lack of energy conservation of the Boris pusher is not the primary factor causing the shortfall in the maximum energy gain.

The bottom panel in Fig.~\ref{fig:energyconservation_nosub} shows the difference between the change in kinetic energy over one time-step, $\Delta \varepsilon_k$, and and the work done over one time-step, $\Delta \varepsilon_w$, in the same simulation. The difference is normalized by $\Delta \varepsilon_t = |e| c E_0 \Delta t$, which is an upper limit for the work done over time interval $\Delta t$. The spikes correspond to stopping points. The spikes point in both directions for each stopping point, which explains why there is not much cumulative error as seen in the top panel of Fig.~\ref{fig:energyconservation_nosub}. 


\begin{figure}
    \centering
    \includegraphics[width=\columnwidth]{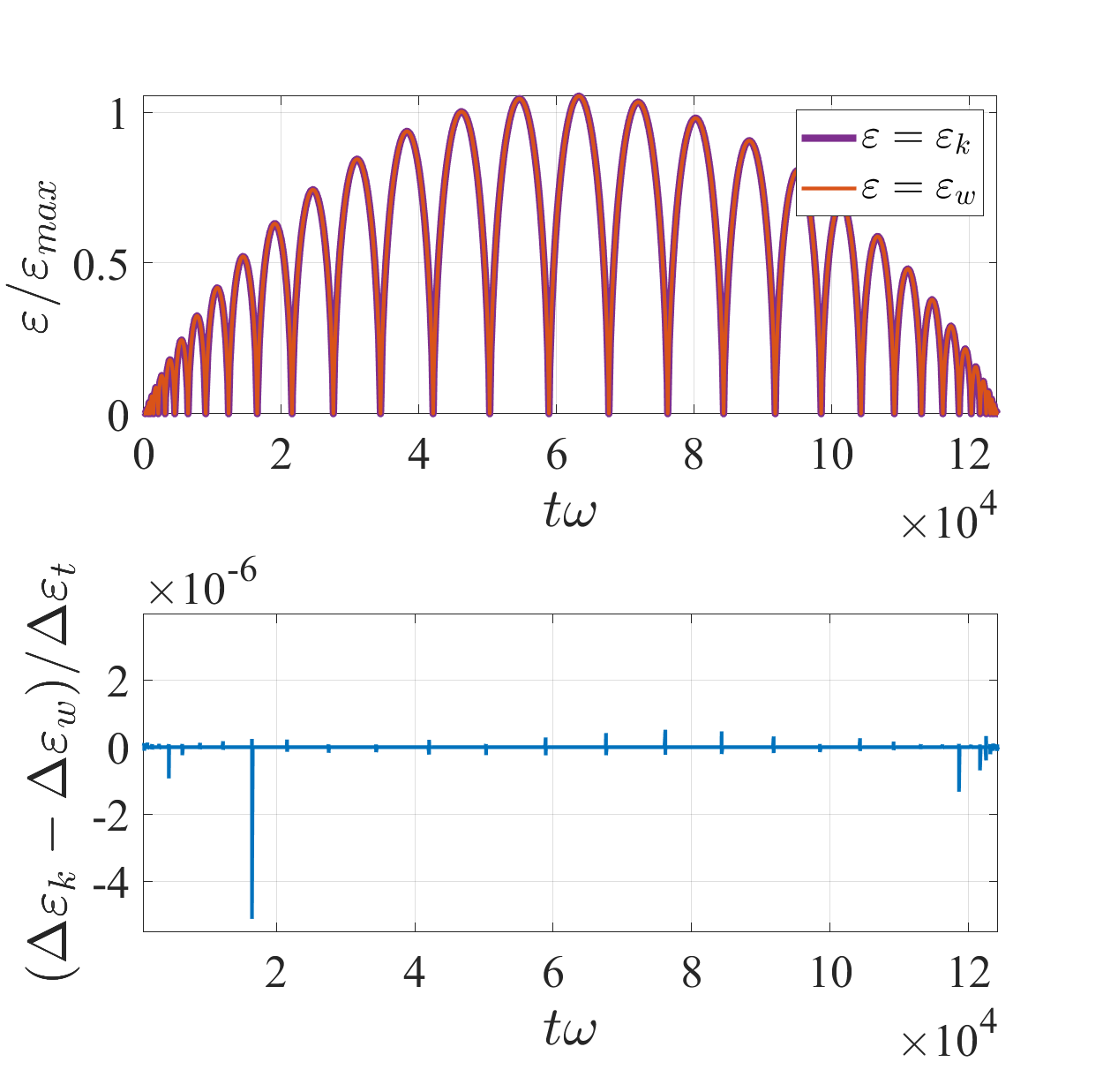}
    \caption{Energy conservation in a PIC algorithm with $3^{rd}$ order temporal interpolation with the sub-cycling algorithm. (top panel) The kinetic energy, $\varepsilon_{k}$, and the net work done, $\varepsilon_{w}$, normalized by $\varepsilon_{max}$. (bottom panel) The difference between the work done and the change in kinetic energy over one $\Delta t$ time-step, $\Delta \varepsilon_{k} - \Delta \varepsilon_{w}$, normalized by $\Delta \varepsilon_{t}$. The simulation parameters are set to $a_0=100$, $\lambda/(\Delta t c) = 75$, $c \Delta t / \Delta x  = 0.99$, and $\Psi_{max}=0.01$. }
    \label{fig:energyconservation_sub}
\end{figure}

In Fig.~\ref{fig:energyconservation_sub}, we turn on the sub-cycling. Comparing the top panels of Figs.~\ref{fig:energyconservation_nosub} and \ref{fig:energyconservation_sub}, we see that the sub-cycling allows the algorithm to better reproduce the theoretically predicted energy gain. The energy increase compared to that in Fig.~\ref{fig:energyconservation_nosub} is the reason for a noticeably longer simulation. The bottom panel shows that the spikes in energy conservation error are improved by many orders of magnitude, as a result of sub-cycling applied near the stopping points. This result is consistent with our general observation that the stopping points are critical parts of the electron trajectory that cause significant errors in electron dynamics.


\end{document}